# INSTRUMENTATION, FIELD NETWORK AND PROCESS AUTOMATION FOR THE CRYOGENIC SYSTEM OF THE LHC TEST STRING

T. Bager, Ch. Balle, E. Blanco, J. Casas, P. Gomes, L. Serio, A. Suraci, N. Vauthier,
LHC Division, CERN 1211 Geneva 23, Switzerland
S. Pelletier, Soteb National Elektro, 60 rue Clement Ader, 01630 St Genis-Pouilly, France

Abstract

CERN is now setting up String 2 [1], a full-size prototype of a regular cell of the LHC arc. It is composed of two quadrupole, six dipole magnets, and a separate cryogenic distribution line (QRL) for the supply and recovery of the cryogen. An electrical feed box (DFB), with up to 38 High Temperature Superconducting (HTS) leads, powers the magnets.

About 700 sensors and actuators are distributed along four Profibus DP and two Profibus PA field buses. The process automation is handled by two controllers, running 126 Closed Control Loops (CCL).

This paper describes the cryogenic control system, associated instrumentation, and their commissioning.

## 1 LAYOUT

String 2 consists of two consecutive LHC half-cells, comprising: two quadrupoles, six 15 m dipoles, their multipole correctors, and the QRL. The chain of magnets is terminated, on the upstream end, by the DFB and, on the downstream end, by the Magnet Return-Box (MRB). The QRL is terminated by its feed and return boxes (QLISF and QLISR).

The DFB houses 3 pairs of 13 kA and 16 pairs of 600 A current leads, to feed 11 electrical circuits. The MRB contains the short circuits and the connection to the QRL, simulating the jumper of the following cell. A 6 kW/4.5 K refrigerator supplies and recovers the cryogen. Fifteen power converters supply the magnets' electrical circuits.

String 2 has a length of about 120m and follows the tunnel curvature of the LHC machine (Figure 1). Most of its elements are heavily instrumented for the experimental program. String 2 is being assembled in two phases. Phase 1 started to be commissioned in April 2001; it comprises all elements except the three dipoles of the second half-cell, which will be added in early 2002.

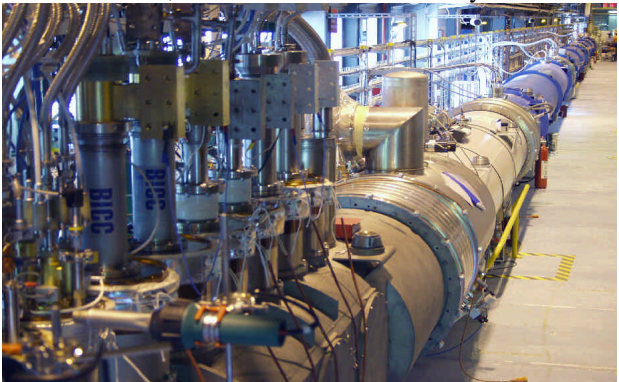

Figure 1: overview of String 2

## 2 PROCESS CONTROL

### 2.1 Structure

The String 2 cryogenic process [2] is handled by two industrial Programmable Logic Controllers (PLC), from Siemens (CPU-S7-414-3DP). One of them is dedicated to the magnets and QRL cryogenics (PLC1), whereas the other controls the DFB cryogenics (PLC2).

The refrigerator and its coupling to String 2 are controlled by an ABB based system, not reported here.

An Engineering WorkStation (EWS) can be remotely connected to the PLCs, through Profibus DP or Ethernet. It is used to program the PLCs and to directly access intelligent devices, for configuration, parameterization and monitoring.

The cryogenic operation and data storage are performed on an Operating WorkStation (OWS), linked to the PLCs through Ethernet. It runs the Supervisory Control And Data Acquisition (SCADA), which is based on PcVue32 and was developed by BARC, India, and by the CERN/LHC/IAS group.

Transients on some process variables are recorded by a fast VME + LabView data acquisition system, developed by INCAA, Netherlands, and by IAS group.

### 2.2 Main Program

Every 100 ms, the main program of each PLC checks if any of the following routines is scheduled for execution (Table 1): CCL, input acquisition, phase sequencer, interlocks, alarms and communication. The time-stamp associated to every call is available to the OWS.

Table 1: instrumentation and routine count per PLC

|  | PLC1 | PLC2 | total |
|---|---|---|---|
| process inputs | 326 | 290 | 616 |
| process outputs | 39 | 42 | 81 |
| total I/O | 365 | 332 | 697 |
| CCLs | 32+7x2=46 | 4+38x2=80 | 126 |
| alarms always valid | 16 | 5 | 21 |
| alarms for specific phases | 25 | 5 | 30 |
| interlocks | 27 | 6+38x2=82 | 109 |

### 2.3 Inputs, Outputs and CCLs

Every one of the 81 process outputs (actuators) is handled by a CCL routine, which implements a PID algorithm. Half of the actuators have a double PID, whereby one out of two process inputs can be selected on the OWS. This leads to 126 PIDs.

The process input to a PID can be filtered (first-order low-pass) or substituted by a constant value. Two setpoints, and their respective offsets, are selectable for each PID; an instantaneous modification of the setpoint value can be limited in slope.

A PID can be in one of the following modes, by order of increasing priority: automatic, forced by process, manual or interlocked. The PID output is respectively: calculated, forced by the process conditions, set by the operator, or driven into a safe position.

CCLs are scheduled for execution with a rate (up to 1Hz) fixed according to the process characteristics. Some heaters have their average power regulated by a Pulse-Width Modulation technique (PWM). Their CCLs are calculated every 10 s and their digital output updated every 100 ms, leading to a 1% resolution.

The about 500 inputs not used in CCLs are read by the acquisition routines and made available in the OWS, for monitoring and diagnostics.

### 2.4 Process Phases

A process phase sequencer tests the start/stop conditions for every operational phase, such as cool-down, normal operation, magnet powering, quench recovery and warm-up. Associated to each phase, automatic actions are programmed for each actuator: choose each of two PIDs, choose each of two setpoints, regulate or force to close or to open, with or without ramp. These automatic actions can be overridden by manual or interlocked modes.

### 2.5 Interlocks

About 30 general interlock conditions are available, in order to force actuators into a safe position, like:
- close valves when liquid level or pressure are high;
- power-off level indicators if no liquid is present;
- stop heaters if a risk of thermal runaway exists.

Moreover, 80 interlocks protect the HTS current leads against excessive cooling: each valve controlling the cooling of the current leads is interlocked with two temperatures at the warm end of the lead.

The interlock thresholds are editable on the OWS and the associated process values and status bit are displayed. This status bit is also visible on each actuator's PID faceplate. Interlocks override automatic and manual actions and they cannot be disabled.

For high-power heaters (>1 kW), local hardware interlocks are activated by the actual temperature of each heating element, independently from the PLC.

### 2.6 Alarms

About 20 alarm conditions are always tested, while 30 others are valid only in specific phases of the process. As for the interlocks, the threshold for each alarm condition is editable on the OWS and the process value and status bit are displayed. Two enable bits are available for alarm activation and for outgoing signal generation.

If any of the enabled alarms lasts more than 1 minute without being acknowledged, or if the communication with a PLC is idle for more than 5 minutes, the OWS automatically warns the operator. A message with the type of alarm is sent through email or cellular phone.

Additionally, when any of the enabled alarms lasts more than 1 minute, or if either PLCs stops, a global hardwired signal is sent to the CERN Technical Control Room (TCR). The responsible person will then be called after a given time. This global signal can also be enabled/disabled/forced by software and hardware.

### 2.7 Communication

Specific routines exchange information between both PLCs and receive data from other systems, such as the refrigerator and the magnet power converter.

On the other hand, critical signals are hardwired:
- status of refrigerator, vacuum and quench detection;
- interlocks to refrigerator and power converters;
- alarm to the TCR.

## 3 FIELDBUS NETWORKS

Instrumentation is spread along the string over 120 m. Most of the process inputs and outputs are accessed by the PLCs through various fieldbus segments (Figure 2).

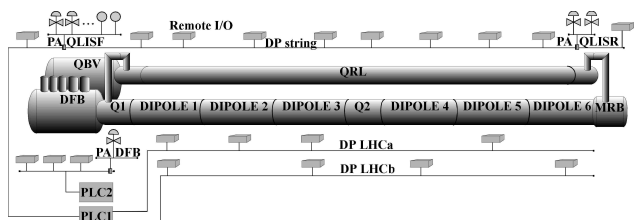

Figure 2: Layout and fieldbus networks

Intelligent devices are directly connected to Profibus PA, operating at 31.25 kbit/s. Conventional devices are connected to analog or digital modules on remote I/O stations (ET200M® from Siemens), which communicate on Profibus DP, running at 1.5 Mbit/s.

### 3.1 Network PLC1

To cover the instrumentation distributed along the magnets and the QRL, three DP and two PA segments have been used.

The sensors that will be used in the LHC machine are distributed over two DP segments, each of them 150 m long. If one of these is down, the sensors on the other will still be enough to run the machine. Each segment comprises 4 remote I/O stations, housed in crates under each dipole, as for the LHC machine.

The third DP segment, extending for about 250 m, gathers instrumentation specific to the experimental program of String 2. Sixteen I/O stations are accommodated in ten racks behind the QRL, whereas 8 other are directly hooked on the QRL. Also on this DP segment are the two PA segments; they include 7 intelligent valves and 3 pressure sensors, installed at one end of the QRL, and 2 valves at the other end. The PLC and some I/O stations are installed in two racks.

### 3.2 Network PLC2

Most of the DFB instrumentation is grouped on a single and short DP segment: 8 I/O stations and a PA segment that handles one intelligent valve positioner.

The 38 valves, on the cooling inlets of the current leads, are driven by local PLC output modules. All the I/O stations and the PLC are installed in two racks. This configuration is representative of the LHC machine, where the PLC and I/O stations of the DFB will be clustered in radiation protected areas.

# 4 INSTRUMENTATION

## 4.1 Temperature

More than 400 resistive temperature sensors, of three types, are used in String 2 (Table 2). As all of them are non-linear, the measured resistance is converted into temperature by an interpolation table. Each Cernox™ and Carbon have a specific table, whereas all Platinum have the same one.

Table 2: temperature sensors

| Type | Quantity | range covered |
|---|---|---|
| Platinum | 295 | 300 K … 30 K |
| Cernox™ | 115 | 300 K … 1.6 K |
| Carbon | 22 | 100 K … 1.6 K |

In general, Platinum thermometers are directly read by analog input modules as a four-wire resistor, without any intermediate signal conditioning.

Most of the Cernox™ and Carbon thermometers are read through linear multi-range signal conditioners (STMS), which produce 4-20 mA magnitude and two bits indicating the range. Logarithmic signal conditioners (LOG) are used for sampling rates faster than a few seconds and have a single 4-20 mA output. Both STMS and LOG are in-house developments [3].

In the current leads, galvanic isolation up to 5 kV is provided by:
- IPAC® conditioners for 190 Platinum thermometers
- STMS conditioners and fuses for 6 Cernox™.

## 4.2 Pressure

On pipes and vessels, absolute pressures up to 100 mbar or to 25 bar are measured by warm transducers, with integrated electronics; three of them operate on Profibus PA whereas 16 other have 4-20mA outputs. The cold mass pressures are measured by 11 passive low-voltage sensors, immersed in liquid helium; their signal conditioners have 4-20mA outputs.

## 4.3 Liquid Level

The eight liquid helium levels inside the vessels and in the superfluid heat exchanger are measured with superconducting-wire gauges (Twickenham Scientific Instruments). To minimize self-heating, the sensors are read in 3s pulsed-mode. The output is 4-20 mA.

## 4.4 Flow

Two mass-flow controllers, ten mass-flow meters and two cryogenic Coriolis mass flow-meters are used to control flow rates and to assess thermal performances. They are read via 4-20 mA current-loops.

## 4.5 Heaters

Electrical heaters are used to warm-up cold gaseous helium leaving String 2, and to evaporate liquid helium in phase separators and in the magnet cold masses.

Twenty-one low-power heaters are driven by level modulated DC voltage. Five higher power heaters, up to 64 kW, are supplied through 220 VAC solid state relays, via the PWM technique. Most of these are in series with programmable relays (Sineax), activated according to the temperature of the respective heating element.

## 4.6 Valves

Ten pneumatic valves with intelligent positioners (Sipart® PS2, from Siemens) are spread on the two Profibus PA segments. Thirty-four classical analog valves are driven with 4-20 mA. Five on/off valves are driven by 0-24 V digital outputs.

# 5 COMMISSIONING

Starting in April 2001, most of the instrumentation and control was checked while the last components were being assembled.

Every instrument had its signal chain verified from the machine to the PLC and to the SCADA. This included cabling, signal processing, coherence between PLC and SCADA databases, and synoptic view. Phases, interlocks and alarms were tested by changing thresholds, while the actuators were inhibited.

Before the final closure of the vacuum vessel, several problems were identified. They were repaired whenever the instrument or cables were still accessible. The origin of these issues has been traced to:
- sensor name and serial number mismatch;
- installation procedure not followed;
- broken wires, sensors failure.

All instruments required for the cooling loops were available when cool-down from 300 K to 4.5 K started, on August 7. The process was gradually set into automatic mode, and the machine could be left unattended.

Apart from the unavoidable time spent for trouble-shooting the components and instrumentation not operating as expected, the overall cool-down from 300 K to 1.9 K took just over 2 weeks, as expected. Once at nominal temperature (1.9 K), the dipole and quadrupole circuits were powered to nominal current (11 850 A). Only then, the control loops were fine-tuned, and cold-mass thermometers and pressure sensors were verified for high accuracy.

During cool-down, the analysis of temperature profiles showed additional inversions and identification errors, mainly on the current leads thermometers.

# 6 CONCLUSION

The installation and commissioning of String 2 cryogenic instrumentation and control was very educative, in view of the methods to be implemented for LHC. The structure of software and hardware documentation was enhanced. Software routines were improved. In the future, sensors installation procedures must be carefully respected, in terms of materials, positioning, and follow-up documentation. In this way, precious time can be saved.

We wish to express our acknowledgement to the operators team, who gave us valuable help during the commissioning of instruments and programs.

# REFERENCES

[1] F. Bordry et al., "The LHC Test String 2: Design Study", LHC Project Report, March 1998.

[2] L. Serio, "The Process of the Cryogenic System for String 2: Functional Analysis", LHC-XMS-ES-0004, LHC Engeneering Specification, Feb 2001.

[3] J. Casas et al., "Signal Conditioning for Cryogenic Thermometry in the LHC", CEC/IMC99, Montreal, Canada, July 1999.